\documentclass[12pt]{article}
\pdfoutput=1
\usepackage{amsfonts}
\usepackage{bbold}
\usepackage{amsmath}
\usepackage{url}
\usepackage{hyperref}
\usepackage{slashed}
\usepackage{tikz}
\usetikzlibrary{decorations.pathmorphing}
\usetikzlibrary{matrix}
\usepackage{graphicx}
\usepackage{epstopdf}
\usepackage{subfigure}
\usepackage{pgfplots}
\usepackage{amssymb}
\usepackage{color}
\usepackage{mathrsfs}
\usepackage{fancybox}
\usepackage{lipsum}
\usepackage{eurosym}
\usepackage{tcolorbox}

\usepackage{tikz,pgfplots}
\usetikzlibrary{decorations.pathmorphing}
\tikzset{snake it/.style={decorate, decoration=snake}}
\usepackage{simpler-wick}
\def\spc{\hspace{1pt}}

\usetikzlibrary{trees}
\usetikzlibrary{shapes.misc}
\usepackage{tkz-euclide}

\def\({\left (}
\def\){\right )}
\def\[{\left [}
\def\]{\right ]}

\numberwithin{equation}{section}

\usepackage{environ}
\NewEnviron{myequation}{%
\begin{eqnarray}
\scalebox{1.05}{$\BODY$}
\end{eqnarray}
}

\newcommand{\beq}{\begin{equation}}
\newcommand{\eeq}{\end{equation}}
 
\newcommand{\bea}{\begin{eqnarray}}
\newcommand{\ea}{\end{eqnarray}}
\newcommand{\barr}{\!\begin{array}}
\newcommand{\earr}{\end{array}\!}

\usepackage{mciteplus}

\def\spc{\hspace{1pt}}

\counterwithout{equation}{section}

\def\smpc{{\hspace{.5pt}}}

\def\calO{{b}}
\def\be{\begin{equation}}
\def\ee{\end{equation}}

\def\la{\langle}
\def\bea{\begin{eqnarray}}
\def\eea{\end{eqnarray}}
\def\is{\!  & \!  = \!  &  \!}
\def\ra{\rangle}

\def\ea{\eea}

\def\be{\bea}
\def\ee{\eea}

\def\ra{\rangle}
\def\la{\langle}

\def\tr{{\rm tr}}

\def\tr{{\rm tr}}
\renewcommand\Large{\fontsize{13.5}{14}\selectfont}
\renewcommand\large{\fontsize{13}{13.5}\selectfont}

\def\spc{\hspace{1pt}}
\def\is{\! &  \! = \! & \!}

\def\darkgreen{green!50!black}
\def\aalpha{\psi}

\def\hhobs{x}
\def\eobs{E}

\begin{document}

\begin{titlepage}

\setcounter{page}{1} \baselineskip=15.5pt \thispagestyle{empty}

\def\hhobs{x}
\def\eobs{E}

${}$
\vspace{1cm}

\begin{center}
{\Large \bf {A microscopic model of de Sitter spacetime with an observer}}\\[1.5cm]
{Damiano Tietto and Herman Verlinde}\\[5mm]
\normalsize{Department of Physics, Princeton University, Princeton, NJ 08544, USA}\\[1cm]
\textbf{Abstract}

\bigskip

\parbox{15cm}{\noindent
We introduce a simple microscopic quantum mechanical model of low-dimensional de Sitter holography with an observer. Using semiclassical gravity and elementary thermodynamic considerations, we derive a formula for the total entropy of a 3D Schwarzschild-de Sitter universe with an observer. We then match this entropy formula with the exactly known spectral density of the double scaled SYK model. Our result gives a de Sitter interpretation of the appearance of two notions of temperature in DSSYK.}

\end{center}

 \vspace{0.3cm}

\vfil
\begin{flushleft}

\today
\end{flushleft}

\end{titlepage}

\def\calO{{b}}
\def\be{\begin{equation}}
\def\ee{\end{equation}}

\tableofcontents

\bigskip
\bigskip

\section{Introduction}

\vspace{-1mm}

\noindent
The Gibbons-Hawking formula \cite{Gibbons:1976ue}\cite{Gibbons:1977mu} for the entropy associated to a static patch provides the most compelling guiding principle for de Sitter holography. Deriving this formula from first principles poses a stringent test for any proposed microscopic model. 

A promising approach to de Sitter holography considers the operator algebra accessible to a local observer 
\cite{Chandrasekaran:2022cip}. This leads one to consider the entangled state of the observer and the ambient spacetime. In this setting, there are at least two different notions of entropy: (1) the total entropy of the combined state of the de Sitter spacetime and the observer,~or (2) the entropy associated with the ambient de Sitter spacetime only. A microscopic holographic model of de Sitter should be able incorporate both notions of entropy and quantify the relation between them.

The high temperature limit of double scaled SYK (DSSYK) has been proposed as a candidate model of low-dimensional de Sitter holography \cite{Narovlansky:2023lfz}
\cite{Susskind:2021esx}
\cite{Rahman:2022jsf}. Elements of the dictionary are emerging \cite{Verlinde:2024znh}\cite{Verlinde:2024zrh} but a full correspondence is still lacking. One immediate task is to establish a clear match between the quantum statistical properties on both sides of the duality.  In this note, we will present a concrete dual interpretation of the DSSYK spectral density and thermodynamics in terms of the above two notions of de Sitter entropy with an observer. Our reasoning relies only on semiclassical gravity and basic elements of quantum statistical physics. 
We start on the gravity side.

\section{3D Schwarzschild-de Sitter thermodynamics}
\vspace{-1mm}

\noindent
In this section we describe the euclidean 3D Schwarzschild-de Sitter (SdS) spacetime and summarize its thermodynamical properties, following the classic work of Gibbons and Hawking \cite{Gibbons:1976ue}\cite{Gibbons:1977mu}. Our discussion will be semiclassical. 

3D SdS spacetime in static coordinates is described by the metric 
\bea \label{esdsone}
    ds^2 \is  \left(r_+^2- r^2 \right ) d\tau^2 + \frac{dr^2}{\left(r_+^2- r^2  \right )}+r^2 d\varphi^2 \; ,\
\eea
Here we use de Sitter units $R_{\rm dS}=1$. The static patch is covered by the coordinate range
\bea \label{esdsonep}
    r \in \left[0, r_+ \right] \, , \hspace{1.cm} \varphi \sim \varphi + 2 \pi \, , \hspace{1.cm} \tau \sim  \tau + \frac{2 \pi}{r_+} \, .\
\eea
The parameter $r_+$ is related to the energy parameter $E$ of the SdS spacetime via  \cite{Spradlin:2001pw}
\bea
\label{edef}
r_+^2 = 1- 8G_N E\, .
\eea
Indeed, a closer examination of the region near $r=0$ reveals that the SdS spacetime contains a conical singularity
\footnote{To make the presence of the matter source more evident, note that we can rewrite \eqref{esdsone} in terms of rescaled coordinates $\bar{r}= r/r_+$, $\bar{\tau} = r_+ \tau$ and $\bar\varphi =r_+\varphi$ as
\bea
\label{esdstwo}
d\bar{s}^2 \is (1-\bar{r}^2) d\bar{\tau}^2 + \frac{d\bar{r}^2}{1-\bar{r}^2} + \bar{r}^2 d\bar{\varphi}^2,\qquad \quad 
\begin{array}{c}{\bar{\varphi}\simeq \bar{\varphi} + 2\pi -\aalpha}\\[2mm]{
\bar{\tau} \simeq \bar{\tau}+2\pi .}\end{array}
\eea} at the location $r=0$. This conical singularity is created by a localized matter source with total energy $E$ related to the deficit angle $\psi$ via
\bea
\label{epsirel}
2\pi - \aalpha = 2\pi r_+ = 2\pi \sqrt{1-8G_NE}.
\eea
For later use, we note that the inverse relation between $\psi$ and $E$ reads
\bea
\label{epsi}
2\pi E(\psi)  = \frac{1}{16\pi G_N} \bigl(4\pi \aalpha  - \aalpha^2.
\bigr).
\eea
This formula will play a key role in what follows. In Appendix A we summarize the derivation of this relation and clarify the physical interpretation of the parameter $E$ with the total energy of the matter source including its gravitational self-energy.

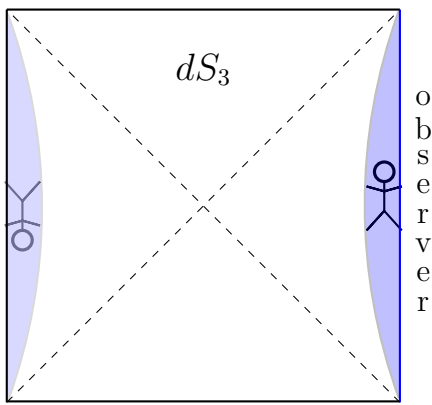
\begin{figure}[t]
\centering
\begin{tikzpicture}[scale=1.3]
\tikzset{
    partial ellipse/.style args={#1:#2:#3}{
        insert path={+ (#1:#3) arc (#1:#2:#3)}
    }
}   
\draw (0,1.4) node {\Large $dS_3$};
\draw[white] (-2.25,1.1) node {o};
\draw (2.25,1.1) node {o};
\draw (2.25,.8) node {b};
\draw (2.25,.5) node {s};
\draw (2.25,.2) node {e};
\draw (2.25,-.1) node {r};
\draw (2.25,-.4) node {v};
\draw (2.25,-.7) node {e};
\draw (2.25,-1) node {r};
\draw[thick] (-2,-2) -- (2,-2);
\draw[very thick] (1.85,.35) [partial ellipse=0:360:.1cm and .1cm];
\draw[thick] (1.85,.25) -- (1.85,.15)--(1.67,.2) -- (1.85,.15)--(2.03,.2) -- (1.85,.15)--(1.85,-.05)--(1.67,-.25) -- (1.85,-.05)--(2.03,-.25) ;
\draw[thick,gray] (-1.85,-.25) -- (-1.85,-.15)--(-1.67,-.2) -- (-1.85,-.15)--(-2.03,-.2) -- (-1.85,-.15)--(-1.85,.05)--(-1.67,.25) -- (-1.85,.05)--(-2.03,.25) ;
\draw[very thick,gray] (-1.85,-.35) [partial ellipse=0:360:.1cm and .1cm];
\draw[thick] (-2.01,2.01) -- (2.01,2.01);
\draw[dashed] (-2.01,-2.01) -- (2.01,2.01);
\draw[thick,blue] (2.01,-2.01) -- (2.01,2.01);
\draw[thick] (-2.01,-2.01) -- (-2.01,2.01);
\draw[dashed] (2.01,-2.01) -- (-2.01,2.01);
\draw[thick,black,fill=blue,opacity=.25] (2,2) arc (160:200:5.8cm); 
\draw[thick,black,fill=blue,opacity=.15] (-2,-2) arc (-20:20:5.8cm); 
\end{tikzpicture}
\caption{We consider a Schwarzschild-de Sitter spacetime with an observer located at the center. The total entropy of the combined system is the sum of the entropies of the SdS spacetime and of the observer.}
\end{figure}

The celebrated Gibbons-Hawking formula equates the entropy of the SdS static patch with $1/4$ of the total length of the observer horizon in Planck units \cite{Gibbons:1977mu}
\bea
\label{sgh}
S_{\rm GH} = \frac{2\pi - \aalpha}{4 G_N}  = \frac{ 2\pi }{4G_N} \sqrt{1- 8G_N E} .
\eea
It is natural to assume that $E$ represents the total energy associated with the static patch spacetime. A naive application of the first law of thermodynamics then identifies the inverse temperature of the static patch~as (note the minus sign)
\bea
\label{betagh}
\beta_{\rm GH} = -\frac{dS_{\rm GH}} {dE} = \frac{2\pi} {\sqrt{1- 8 G_N E}}.
\eea
This temperature should be interpreted as governing the first law of thermodynamics
\bea
\label{ghfirst}
\Delta S_{\rm GH} \is -\beta_{\rm GH} \Delta E
\eea
for processes that involve the exchange of energy between the interior and the exterior of the static patch, i.e. that involve energy transport through the observer horizon and a resulting change in the total horizon length.

Now suppose we place a local observer at the origin $r=0$, as depicted in figure 1. The total energy of the observer equals $E_\text{obs}$.  In general, $E_{\rm obs}$ does not need to equal the total energy $E(\psi)$ that creates the conical deficit $\psi$. However, it is reasonable to make the physical assumption that the observer energy is bounded by
\bea
\label{eless}
0\leq E_{\rm obs} \leq E(\psi).
\eea
In the semiclassical limit, the observer contribution to the euclidean partition function is
\bea
\label{zetobs}
    \mathcal{Z}_\text{obs} =  e^{-  I_\text{obs}} \qquad I_{\rm obs} = E_{\rm obs} \oint d\tau_\text{prop}\,.
\eea
Here $\oint d\tau_\text{prop}$ represents the total euclidean proper time of circular worldline trajectory of the observer. From \eqref{esdsone}-\eqref{esdsonep} we read off that proper time measured by the observer is 
\bea \label{observer proper time}
    \oint d \tau_\text{prop} = \oint \sqrt{ds^2} \is 2 \pi \, 
\eea
in de Sitter units. It is natural to conclude~from this that the local observer at $r=0$ experiences the ambient SdS spacetime as a thermal bath with constant inverse temperature
\bea
\label{betads}
\beta_{\rm dS} = 2\pi,
\eea
independent of the value of the energy $E_{\rm obs}$ or deficit angle $\aalpha$. 
Note that this inverse temperature \eqref{betads} differs from the GH temperature \eqref{betagh}.

Concretely, equation \eqref{betads} means that if the observer would extract an energy amount $\Delta E_{\rm obs}$ from the ambient SdS spacetime, they would conclude that the entropy $S_{\rm SdS}$ of the ambient spacetime is reduced by an amount
\bea
\label{deltas}
\Delta S_{\rm SdS} = -\beta_{\rm dS} \Delta E_{\rm obs}\, .
\eea
This equation is physically different from the first law \eqref{ghfirst} for the GH entropy. Instead, \eqref{deltas} represents the first law for processes that involve energy exchange between the ambient spacetime and the observer. If this process takes place locally, it does not change the deficit angle $\psi$ nor the GH entropy $S_{\rm GH}$.

In the remainder of this paper, we will develop a microscopic quantum statistical description of the de Sitter static patch and the observer that unifies the GH entropy formula \eqref{sgh}  with the fact that the observer experiences a constant inverse temperature \eqref{betads}. Before turning to this discussion, we first summarize the quantum statistical and thermodynamic properties of our proposed dual system.

\section{DSSYK partition function and thermodynamics}
\vspace{-1mm}

\noindent
The SYK model consists of $N$ majorana oscillators $\{\chi_i,\chi_j\}=2\delta_{ij}$ with a $p$-body interaction term 
with gaussian random couplings
\bea
H_{{\!}_{\rm \spc SYK} } \! \is  \!  \; i^{p/2} \sum_{i_1\ldots i_p} J_{i_1\ldots i_p} \chi_{i_1}\ldots \chi_{i_p}\, .
\eea 
This model is exactly soluble in the scaling limit 
\cite{Berkooz:2018jqr}
\bea
\begin{array}{c}{N\to \infty}\\[.5mm] { p \to \infty} \end{array} \qquad \lambda = \frac{2p^2}{N}\ \, {\rm fixed}
\eea
Physical quantities that remain finite in this limit depend only on the dimensionless number $\lambda>0$. In the following, we work in the semiclassical regime $\lambda\ll1$ and choose units so that the coupling ${\cal J}= 1$.

The DSSYK partition function can be written as an integral over an auxiliary variable, which somewhat presumptively we will also denote by $\aalpha$, as 
\bea
\label{zsyk}
    {\cal Z}_{\rm SYK}(\beta)    \is  \int_0^\infty \!\! d{{\cal E}}(\aalpha)\, e^{S_{\rm SYK}(\aalpha) -\beta {\cal E}(\aalpha)} .  
 \eea
Here the energy and entropy are parametrized via 
\bea
{\cal E}(\aalpha) \is  -\frac{2}{\lambda}\spc {\sin( \aalpha/2)} \\[1mm]
S_{\rm\smpc SYK}(\aalpha) \is  {S_0} + \frac{4\pi^2}{\lambda} - \frac{\aalpha^2}{2\lambda},
\label{ssyk}
\eea
where $S_0$ is an overall entropy shift proportional to~$N$.
Equation \eqref{zsyk} is an exact expression. Since the energy ${\cal E}$ is a periodic function in $\aalpha$, it is natural to substitute $\aalpha \to \aalpha + 2\pi n$ and split the integral as a sum over $n$ of a $\aalpha$-integral from $-\pi$ to $\pi$. The spectral measure then turns into a Jacobi $\vartheta$-function 
\cite{Berkooz:2018jqr}. 

We see that the spectral measure of DSSYK behaves as a gaussian with a maximum at $\aalpha=0$.  The inverse temperature
\bea
\beta_{\rm SYK} = \frac{dS_{\rm SYK\!\!}}{d{\cal E}} \; = \frac{\aalpha}{\cos(\aalpha/2)}
\eea
vanishes at the maximum entropy point $\aalpha=0$. This infinite temperature state is conjectured to describe pure de Sitter space.

One of the hallmarks of DSSYK is that thermal correlation functions exhibit a periodicity in euclidean time that differs from the expected KMS period $\beta_{\rm SYK}$ \cite{Maldacena:2016hyu}. Instead, they are periodic with the fake inverse temperature
\bea
\label{betaratio}
\beta_{\rm fake} = \frac{2\pi}{\aalpha} \spc \beta_{\rm SYK} = \frac{2\pi}{\cos(\aalpha/2)}.
\eea
If we would  introduce a corresponding notion of fake entropy $S_{\rm fake}$ such $\beta_{\rm fake} = \frac{dS_{\rm fake}}{d{\cal E}}$, we would find that
\bea
S_{\rm fake} = -\frac{2\pi \aalpha}{\lambda} + {\rm const}.
\eea
Note that the fake inverse temperature $\beta_{\rm fake}$ is finite at $\aalpha=0$ and the fake entropy $S_{\rm fake}$ is linear in the angular variable $\aalpha$. Both properties are suggestive of a relation with de Sitter space. To make this dual interpretation concrete, however, we need to explain on the gravity side why the physical entropy $S_{\rm SYK}(\aalpha)$ takes the form \eqref{ssyk}.

\section{De Sitter thermodynamics with an observer}

\vspace{-1mm}

\noindent
We now return to the gravity side. We like to address two seemingly unrelated questions: 

\bigskip

\noindent
${}$~~~\parbox{15cm}{1. How do we reconcile the constant inverse temperature $\beta_{\rm dS}$ experienced by the observer with~the GH formula for the entropy of the static patch?}

\bigskip

\noindent
${}$~~~\parbox{15cm}{2. How can we explain the DSSYK spectral density \eqref{ssyk} and the appearance of two inverse temperatures $\beta_{\rm SYK}$ and $\beta_{\rm fake}$ from the gravity side?}

\bigskip

\noindent
As we will see, the answers to these two questions are linked and both rely on including the effect of the observer on the entropy of the de Sitter patch.

Let us introduce the Hilbert subspace with given size of the cosmological horizon
\bea
{\cal H}^{\rm max}_{\rm \psi}\is \begin{array}{c}{\mbox{Hilbert space of the static SdS patch}}\\ {\mbox{with deficit angle $\psi$ plus the observer}}
\end{array}\nonumber\
\eea
We denote the dimension of this Hilbert space by
\bea
N_{\rm \aalpha} = {\rm dim}{\cal H}^{\rm max}_\aalpha  = e^{S_{\rm max}(\aalpha)}.
\eea
Hence $S_{\rm max}(\aalpha)$ is the microscopic entropy of the combined system of the de Sitter static patch with given deficit angle  $\aalpha$ and the observer. We will assume that the combined system is in the micro-canonical maximum entropy state 
\bea
\label{rhomax}
\rho^{\,\rm max}_\aalpha = \frac{1}{\! N_{\rm \aalpha}\!\!}\; \mathbb{1}_\aalpha =  \frac{1}{\! N_{\rm \aalpha}\!\!}\;\sum_{|i\ra \in {\cal H}_\aalpha} |\spc i \spc \ra \la \spc i \spc |\, 
\eea
with entropy $S_{\rm max}(\aalpha)$. 
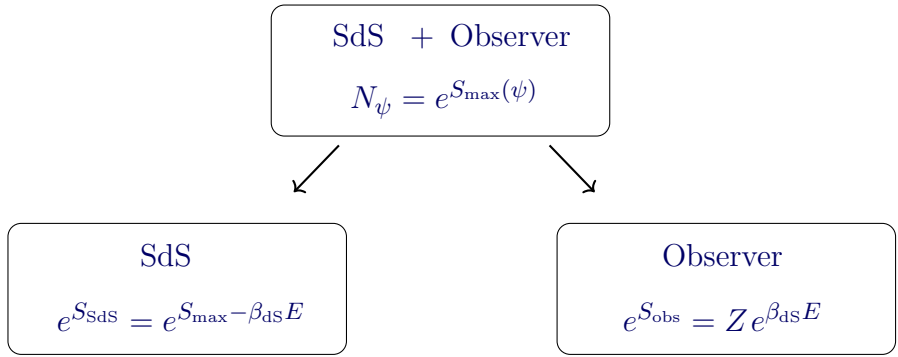
\begin{figure}[t]
\centering
\begin{tikzpicture}[yscale=-.83,xscale=1]    
       \draw[rounded corners=5pt] (1.5,2.5)  rectangle ++(4.5,2.1);
       \draw[rounded corners=5pt] (-2.3,-1)  rectangle ++(4.5,2.1);
       \draw[rounded corners=5pt] (-5.8,2.5)  rectangle ++(4.5,2.1);
        \draw[black,thick, <-] (-2,2) -- (-1.4,1.25) ;
        \draw[black,thick,  <-] (2,2) -- (1.4,1.25) ;
         \node[blue!40!black] at (-3.7,3) {\large SdS};
         \node[blue!40!black] at (-3.5,4) {\large $e^{S_{\rm SdS}}  =  e^{S_{\rm max} -  \beta_{\rm dS} E}$};   
         \node[blue!40!black] at (.1,-0.5) {\large SdS \, +\, Observer};
         \node[blue!40!black] at (0,.5) {\large $N_{\rm \aalpha} = e^{S_{\rm max}(\aalpha)}$};
         \node[blue!40!black] at (3.7,3) {\large Observer};
         \node[blue!40!black] at (3.7,4) {\large $e^{S_{\rm obs}} = Z \spc e^{ \beta_{\rm dS} E}$}; 
\end{tikzpicture}
\caption{We view the combined system of the SdS static patch and the observer as an entangled bound state. The total entropy of the combined system is the sum of the entropies of the two subsystems. }
\end{figure}

We will now determine $S_{\rm max}(\aalpha)$ by applying the holographic postulate and simple quantum statistical reasoning to the combined state. In what follows, we will treat the cosmological deficit angle $\psi$ as a superselection sector that can not be changed by physical operators accessible to a local observer near $r=0$.
We introduce
\bea
{\cal H}^{{}{\rm SdS}}_\psi\, \is \begin{array}{c}{\mbox{Hilbert space of the ambient static }}\\ {\mbox{SdS patch with deficit angle $\psi$}}
\end{array} \nonumber\\[2mm]
{\cal H}_{\rm  obs}\, \is \ {\mbox{Hilbert space of the observer}} \nonumber
\eea

The observer Hilbert space ${\cal H}_{\rm  obs}$ is a direct sum of energy eigenstates with energy $E_{\rm obs}$. Evidently, to fit inside the SdS space with deficit angle $\aalpha$, we must require that $E_{\rm obs}\leq E(\aalpha)$. Likewise, the SdS Hilbert space ${\cal H}^{{}{\rm SdS}}_\psi$ is a direct sum of energy eigenstates with energy $E_{\rm SdS}$. By definition, the energy $E_{\rm SdS}$ of the ambient SdS static patch makes up the difference between the observer energy and the total energy $E_{\rm tot} = E_{\rm SdS} + E_{\rm obs} = E(\psi).$
We parametrize the solutions to this equality via
\bea
\label{eparam}
\begin{array}{cc} {E_{\rm obs} = E\qquad \quad\ }\\[2mm]
 {E_{\rm SdS}  = E(\psi) - E}\end{array} \qquad 0\leq E\leq E(\aalpha) .
\eea
Localized physical observables that act within ${\cal H}^{\rm max}_{\rm \psi}$ can change $E$ but they can not change the value of the cosmic deficit angle $\psi$.

The Hilbert space of the combined system is the subspace of ${\cal H}^{\rm SdS}_\psi \otimes {\cal H}_{\rm obs}$ 
selected by the above total energy condition 
\bea
\quad {\cal H}^{\rm max}_{\rm \psi} = \bigl({\cal H}^{\rm SdS}_\psi \otimes {\cal H}_{\rm obs}\bigr)_{E_{\rm tot} = E(\aalpha)}.
\eea
This space is spanned by entangled states of the form
\bea
\label{embed}
| \spc i \spc \ra_{\rm max} \is \sum_{j,n} C^{\spc i}_{jn} \, | j\ra_{{\!}_{\rm SdS}}|n\ra_{{\!}_{\rm obs}},
\eea
with $| \spc i \spc \ra_{\rm max}$, $| j\ra_{{\!}_{\rm SdS}}$ and $|n\ra_{{\!}_{\rm obs}}$ some arbitrary chosen basis in the respective Hilbert spaces. The coefficients $C^{\spc i}_{jn}$ describe the embedding of the physical Hilbert space ${\cal H}^{\rm max}_{\rm \psi}$ into the tensor product Hilbert space  ${\cal H}^{\rm SdS}_\psi \otimes {\cal H}_{\rm obs}$. They are assumed to form a random tensor ${\bf C}$ compatible with unitarity\footnote{Defining the tensor operator ${\bf C}$ such that $C^i_{jn} = \la j ,  n|\spc {\bf C}\spc | i\ra$,  unitarity and randomness imply that~\cite{Verlinde:2012cy}
\bea
{\bf C}^\dag {\bf C} = \mathbb{1}_\psi^ {\rm max}  \  &; & \ {\bf C}\, {\bf C}^\dag = \bigoplus_E \frac{1}{Z}\; \mathbb{1}^{\rm SdS}_{\aalpha,E} \otimes  \, \mathbb{1}^{\rm obs}_{E}
\eea
The first identity is exact; the second equation is a statistical statement that holds~${O}(e^{-\frac{S_{\rm max}\!\!}{2}})$.}
and the total energy constraint 
\bea
H_{\rm tot} |\smpc i\smpc \ra_{\rm max} = E(\psi)|\smpc i\smpc \ra_{\rm max}.
\eea

Equation \eqref{embed} specifies the embedding of the bound state $| \spc i \spc \ra_{\rm max}$ into the product Hilbert space 
of the SdS patch and the observer.
Applying this embedding map to the maximal entropy state \eqref{rhomax}, combined with some standard statistical reasoning, determines that 
$\rho^{\, \rm max}_\aalpha$ takes the diagonal form (up to off diagonal terms of order $e^{-S_{\rm max}/2}$)
\bea
\label{rhodeco}
\rho^{\, \rm max}_\aalpha \is  \frac{1}{\! Z N_{\aalpha}\!\!}\, \bigoplus_E \; \mathbb{1}^{\rm SdS}_{\aalpha,E} \otimes  \, \mathbb{1}^{\rm obs}_{E}. 
\eea
Here the direct sum over $E$ runs over a dense collection of small energy windows with width $\Delta E\ll 1$ given by the energy resolution of the observer, and
$\mathbb{1}^{\rm obs}_{E}$ and $\mathbb{1}^{\rm SdS}_{\aalpha,E}$ denote projection operators on the respective subspaces with energy $E$.  

Next we use the observational input, derived from the euclidean SdS geometry \eqref{esdsone},  that the ambient SdS spacetime has a constant inverse temperature $\beta_{\rm dS}$. From this we deduce
\bea
\label{sds-spec}
{\rm tr} \, \mathbb{1}^{\rm SdS}_{\aalpha,E} 
\!&\!\equiv \!&\! N_{\aalpha, E} = N_{\aalpha} \, e^{- \beta_{\rm dS} E}\\[2mm]
Z \is \sum_{|E\ra \in {\cal H}_{\rm obs}} e^{-\beta_{\rm dS} E}  \, .
\eea
The formula for $Z$ ensures that $\tr\rho^{\, \rm max}_\aalpha =1$. The quantity $N_{\psi,E}$ represents the total number of states of the ambient static patch within a small energy window centered on $E_{\rm SdS} = E(\psi) - E$. Eqn \eqref{sds-spec} states that the microscopic entropy $S_{\rm SdS}(\aalpha, E) =\log N_{\psi, E}$ of the static patch with deficit angle $\psi$ and energy $E_{\rm SdS} = E(\psi) - E$ equals
\bea
\label{entone}
S_{\rm SdS}(\aalpha, E) \is S_{\rm max}(\aalpha) - \beta_{\rm dS} E\, . \ 
\eea
The linear $E$-dependence reflects the observational fact that the observer experiences a constant de Sitter temperature.

We are now ready to determine $S_{\rm max}(\aalpha)$. Let $S_{\rm SdS}(\aalpha)$ denote the entropy of the ambient  SdS geometry with horizon deficit angle $\psi$ of an observer with the maximal allowed energy $E=E(\aalpha)$. By our holographic postulate, this entropy $S_{\rm SdS}(\aalpha)$ is given by the Gibbons-Hawking formula \eqref{sgh}. Combining this identification with \eqref{entone}, we deduce that
\bea
\label{entwo}
S_{\rm SdS}(\aalpha) = S_{\rm max}(\aalpha) - \beta_{\rm dS} E(\psi) = S_{\rm GH}(\psi)\, .
\eea
 The first equality follows from setting $E=E(\psi)$ in \eqref{entone} while the second equality follows from the holographic postulate. 
Hence we find that the maximum entropy of the ambient de Sitter patch with given deficit angle $\psi$ equals
\bea
\label{ssum}
\boxed{\ S_{\rm max}(\aalpha)  = S_{\rm GH}(\aalpha) + \beta_{\rm dS} E(\aalpha){}_{\strut} \LARGE \strut\,}
\eea 
with $E(\psi)$ the energy \eqref{epsi} required to create the conical deficit angle $\psi$. It is not difficult to see that the maximum entropy $S_{\rm max}(\aalpha)$ is also equal to the entropy of the combined SdS+observer system with deficit angle $\psi$.

Our model allows the observer energy $E_{\rm obs}\!=\!E$ to be less than the total energy $E(\psi)$ needed to create the angle deficit. However, suppose we ask: what is the maximal entropy state consistent with the presence of an observer with energy $E_{\rm obs}=E$? The entropy of this state is given by
\bea
S_{\rm SdS}(E) \! & \! \equiv \! & \! \sup_{{\aalpha}}\;  S_{\rm SdS}(\aalpha,E) 
= S_{\rm GH}(E),
\eea
where the supremum is taken over all deficit angles $\aalpha$ compatible with the inequality $E\leq E(\aalpha)$. The supremum occurs at the value of $\aalpha$ for which the equality $E(\aalpha) = E$~holds.

\section{Matching de~Sitter and DSSYK thermodynamics }

\noindent
Equation \eqref{ssum} is our main result.~In~short, it states that the entropy of the SdS+observer system is given the sum of the GH entropy and the entropy that an observer with maximal energy $E_{\rm obs} = E(\psi)$ has extracted from the ambient spacetime.  Here we will use it to give a de Sitter interpretation of the thermodynamical properties of DSSYK.

\subsection{De~Sitter~interpretation~of~SYK~spectral~density}

\vspace{-1mm}

\noindent
Via the relation \eqref{ssum}, we can now make direct contact with the spectral entropy formula \eqref{ssyk} of the double scaled SYK model. Plugging in the GH formula \eqref{sgh} and the relation \eqref{epsi} for the energy gives 
\bea
S_{\rm max}(\aalpha)  \is
\frac{2\pi-\aalpha}{4G_N} + \frac{4\pi \aalpha - \aalpha^2}{16\pi G_N} \nonumber \\[-2mm]
\label{smax}
\\[-2mm]\nonumber
 \is  \frac{\pi}{2 G_N} -\frac{\aalpha^2}{16\pi G_N}\, .
\eea
This formula should be compared with the expression \eqref{ssyk} for the SYK spectral density.

In conclusion, we have argued that the ambient SdS static patch entropy is given by \eqref{entone}-\eqref{entwo}. These formulas are compatible with the GH entropy formula and the presence of a constant de Sitter temperature.
 Moreover, we have found a natural identification between the DSSYK spectral entropy \eqref{ssyk} and the maximum entropy \eqref{smax} of the combined SdS+observer system
\bea
\quad \ \boxed{\ S_{\rm SYK}(\psi) = S_0 + S_{\rm max}(\psi) {}_{\strut} \LARGE \strut\,}
\eea
up to the overall constant term $S_0$. This identification holds provided we identify the angles $\psi$ on each side and set the dimensionless DSSYK coupling $\lambda = {2p^2}/{N}$ equal to
\bea
\label{lambdaid}
\lambda \is 8\pi G_N.\
\eea

\subsection{De~Sitter~interpretation~of~the~two~temperatures}

\vspace{-1mm}

\noindent
As an added bonus of the above discussion, we can now give a concrete gravity interpretation of the appearance of two different inverse temperatures. 
\bea
\beta_{\rm max} \is -\frac{dS_{\rm max}}{dE} = \frac{\psi}{8\pi G_N} \frac{d\psi}{dE}\, , \ \\[2mm]
\beta_{\rm GH} \is  -\frac{dS_{\rm GH}}{dE} = \frac{1}{4G_N} \frac{d\psi}{dE}\, .\
\eea
Here we used equations \eqref{smax} and \eqref{sgh}. The ratio of these two inverse temperatures 
\bea
\beta_{\rm GH} \is \frac{2\pi}{\psi}\, \beta_{\rm max} \
\eea
matches the ratio \eqref{betaratio} of the physical and fake inverse temperature in DSSYK. In our proposed interpretation, the high temperature limit of the SYK model describes a de Sitter universe {\it including} an observer. The presence of the observer increases the entropy by an amount $\beta_{\rm dS} E(\psi)$ to become equal to $S_{\rm max}(\psi)$ and the combined system heats up to a close-to-infinite temperature state with inverse temperature $\beta_{\rm max} = \frac{\psi}{2\pi} \beta_{\rm dS} = \beta_{\rm SYK}$ . Nevertheless, the correlation functions of physical operators look like thermal correlators with finite inverse temperature $\beta_{\rm GH} = \beta_{\rm fake}$. The appearance of the two temperatures are reconciled by the above physical argument, or in the dual DSSYK model, by explicit computation of the physical correlators.  

As an example, consider the two point-function of physical operators ${\cal O}$ evaluated in a maximal entropy state
\bea
\label{gabtwo}
G(\tau) \!\is\! \bigl\la \Psi_{0} \bigl| \spc {\cal O}(\tau) \spc {\cal O}(0)\spc  \bigr|\Psi_{0} \bigr\ra .\
\eea
Localized physical operators ${\cal O}$ can not modify the conical deficit angle of SdS spacetime. Hence, local physical operators allow for a decomposition 
\bea
\label{physop}
{\cal O}(\tau) \is \int\! dt \, {\cal O}_{\rm SdS}(\tau\!-\!\spc t) {\cal O}_{\rm obs}(t), \
\eea
where ${\cal O}_{\rm obs}$ and ${\cal O}_{\rm SdS}$ are elements of a pair of mutually commuting operator subalgebras ${\cal A}_{\rm obs}$ and ${\cal A}_{\rm SdS}$ of the full SYK operator algebra.  A natural choice is to let ${\cal A}_{\rm obs}$ be generated by a set of local DSSYK scaling operators $O_{\rm \Delta}$ and to choose  ${\cal A}_{\rm SdS}$ to be generated by the mirror (or shadow) operators in the commutant of  ${\cal A}_{\rm obs}$. The space of physical operators \eqref{physop} forms a subalgebra $ ({\cal A}_{\rm obs}\otimes {\cal A}_{\rm SdS})^{H_{\rm tot}}$
of the tensor product algebra selected by the condition that they must commute with the total Hamiltonian $H_{\rm tot} = H_{\rm obs} + H_{\rm SdS}$. 

In this way, we have made contact with the construction of physical operators in  \cite{Narovlansky:2023lfz} in terms of a doubled SYK model. Note, however, that in the above construction, the two algebras are assumed to be commuting subalgebras of a single microscopic SYK model.\footnote{In this interpretation, it is natural to think of the total Hamiltonian $H_{\rm tot}$ as the modular Hamiltonian associated with the maximal entropy state $|\Psi_0\ra$. The above construction also bears a close similarity to the description of a de Sitter observer in \cite{Chandrasekaran:2022cip} in terms of the crossed product of a type III von Neumann algebra.}

In \cite{Narovlansky:2023lfz} it was shown that, in the semiclassical $\lambda \to 0$ limit, the two point function of suitable chosen set of physical DSSYK operators \eqref{physop} reproduces the two-point function along a timelike worldline in 3D de Sitter space. In particular, it exhibits periodicity in euclidean time with period given by the inverse de Sitter temperature.  To understand the microscopic mechanism underlying this result, we can look at the Fourier transform of the two-point function  \eqref{gabtwo}
\bea
G(E) \is  e^{S_{\rm SYK}(E)}| \la E| {\cal O}| \Psi_0\ra|^2 \, .
\eea
One can think of $G(E)$ as the spectral density of the intermediate factorization channel between the insertions of the two physical operators ${\cal O}(\tau)$ on ${\cal O }(0)$. Each operator ${\cal O}$ acts like an random matrix between energy eigenstate similar to the tensor $C^{\spc i}_{jn}$ introduced in equation \eqref{embed}. Using the same statistical reasoning that led to equations \eqref{rhodeco}-\eqref{sds-spec}, or by direct computation in DSSYK  \cite{Narovlansky:2023lfz}, one finds that $G(E)$ exhibits the characteristics of a thermal Green's function with inverse temperature $\beta_{\rm dS}$.

\section{Concluding remarks}

\vspace{-1mm}

We presented a simple microscopic quantum mechanical model of low-dimensional de Sitter holography with an observer. Using semiclassical gravity and elementary thermodynamic considerations, we derived a formula for the total entropy of the combined system of 3D Schwarzschild-de Sitter universe including the observer and found a precise match with the microscopic entropy of DSSYK.

As shown in figure 3, the entropy $S_{\rm max}$ of the combined system of the SdS static patch and the observer is the sum of the Gibbons-Hawking entropy $S_{\rm GH}$  and the entropy $S_{\rm obs}(\psi) = \beta_{\rm dS} E$ that the observer has extracted from the ambient SdS spacetime. The derivation of this result relies on the  observation, inferred from the explicit form of the SdS spacetime metric, that the observer experiences the SdS static patch as a thermal environment with constant inverse temperature $\beta_{\rm dS} = 2\pi$.  Our model incorporates this property by allowing the observer to have arbitrary energy less than the total energy $E(\psi)$ in the SdS spacetime with given deficit angle $\psi$.

We have found that the formula for the entropy $S_{\rm max}(\psi)$ exactly matches the known spectral density of the double scaled SYK model. The match is based on an identification between the spectral angle of DSSYK with the deficit angle of the SdS spacetime. 

We end with some comments and open questions.

\smallskip

\begin{figure}[t]
\centering
\begin{tikzpicture} [xscale=3,yscale=1.8,domain=0:2]
\draw[thick,->] (0, 0) -- (2.3, 0) node[right] {$\aalpha$};
  \draw[thick,->] (0, 0) -- (0, 2.3) node[left] {$S(\aalpha)$};  
   \draw[\darkgreen,->] (.5, 2) -- (.5, 1.75);    
  \draw[\darkgreen,->] (1, 1.95) -- (1, 1.5);    
  \draw[\darkgreen,->] (1.5, 1.87) -- (1.5, 1.25);    
 \draw[\darkgreen,->] (.25, 2) -- (.25, 1.875); 
 \draw[\darkgreen,->] (.75, 1.985) -- (.75, 1.625);  
  \draw[\darkgreen,->] (1.25, 1.92) -- (1.25, 1.375); 
  \draw[\darkgreen,->] (1.75, 1.82) -- (1.75, 1.125);  
  \draw[dashed] (2,0) -- (2, 2.3);
  \draw (2,-.25) node {$\pi$};
    \draw[thick,color=red, domain=0:2]   plot (\x,{2-\x/2}) ;
    \draw[thick,color=blue, domain=0:2]   plot (\x,{2- \x*\x/16}) ;
    \draw[thick,color=\darkgreen, domain=0:2]   plot (\x,{\x/2- \x*\x/16}) ;
\draw[blue] (1.65,2.05)    node {$S_{\rm max}$};
\draw[\darkgreen] (1.65,.38) node {$\beta_{\rm dS} \eobs$} ;
 \draw[red] (1.64,1) node {$S_{\rm GH}$};
 \end{tikzpicture}
 \caption{The entropy $S_{\rm max}(\psi)$ ({\textcolor{blue}{blue}}) of the combined system of the SdS static patch and the observer is the sum of the Gibbons-Hawking entropy $S_{\rm GH}(\psi)$ ({\textcolor{red}{red}})  and the entropy $S_{\rm obs}(\psi) = \beta_{\rm dS} E(\psi)$ ({\textcolor{\darkgreen}{green}})  that the observer of energy $E(\psi)$ has extracted from the ambient SdS spacetime.}
\end{figure}
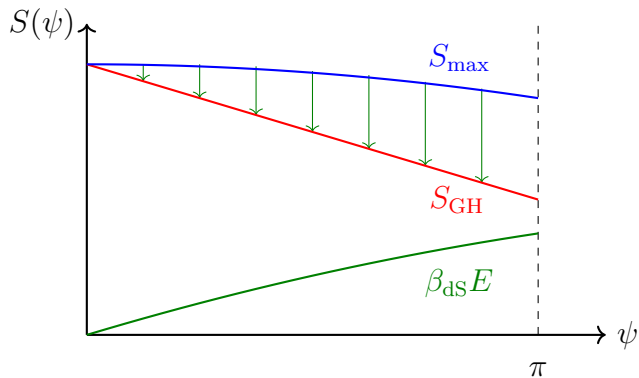

1. Our reasoning gives a direct gravity interpretation of both the physical and fake temperature in DSSYK. We can write two versions of the first law of thermodynamics
\bea
\label{firstlaw}
dS_{\rm max}(\psi) \is  \beta_{\rm SYK}(\psi)\smpc d\smpc{\cal E}_{\rm SYK}(\psi) , \nonumber \\[-1,75mm]\\[-1.75mm]\nonumber
dS_{\rm GH}(\psi) \is \beta_{\rm fake}(\psi) \smpc d\smpc{\cal E}_{\rm SYK}(\psi)  ,
\eea
with ${\cal E}_{\rm SYK}(\psi)$ the DSSYK energy \eqref{eparam}. The right-hand sides in \eqref{firstlaw} are pure DSSYK quantities. Via our  holographic dictionary, we have given a de Sitter gravity interpretation of the two left-hand sides as the change in entropy of the SdS+observer system and of the ambient SdS spacetime, respectively.
 
\smallskip

2. In equation \eqref{zsyk} we have written the DSSYK partition function as an integral over all real values of the angular variable $\psi$. In the SdS geometry, the deficit angle $\psi$ is bound to lie in a finite interval between $0$ and $2\pi$. As mentioned, we can rewrite \eqref{zsyk} as an integral over a finite range of $\psi$ as follows
\bea \label{syk basic partition function}
 {\cal Z}_{\rm SYK}(\beta)  \is  \int_{0}^{2\pi}  \! {d\psi}\, \rho_{\rm SYK} (\psi) \, e^{-\beta {\cal E}(\psi)}\nonumber \\[-3mm]\\[-1mm]\nonumber
  \rho_{\rm SYK} (\psi)  \is e^{S_0}  \cos\bigl(\mbox{\Large ${\frac\psi 2}$}\bigr) \sum_{n \in \mathbb{Z}} (-1)^n e^{-\frac{1}{2\lambda}(\psi + 2\pi n )^2}
\eea
Noting the identification \eqref{lambdaid}, we see that the spectral density $\rho_{\rm SYK}$ takes the form of an infinite sum of saddle point contributions. For $\psi$ between $0$ and $\pi$, the $n=0$ term is dominant saddle point. Our dictionary interprets this saddle point as the Schwarschild-de Sitter spacetime. The other terms are non-perturbative corrections due to formal saddle points with deficit angle $\psi + 2\pi n$. 

The low temperature regime of DSSYK corresponds to the transition point $\psi = \pi$ between the $n=0$ and $n=-1$ saddle points. The interpretation of  low temperature regime as a holographic dual to AdS$_2$ gravity can not easily be understood via a straightforward semi-classical extrapolation of the de Sitter interpretation of the high temperature regime.

\smallskip

3. We have found a new match between the thermodynamic equations of 3D de Sitter space plus an observer and double scaled SYK. This result adds yet another entry to the dual dictionary between the two systems. Other elements of the correspondence include precise equalities between partition functions, low-point correlators, and quantum symmetries. An apparent point of difference between the two systems, however, is that the DSSYK entropy includes an (infinite) constant contribution $S_0$, whereas, at first sight, a dual microscopic model of 3D de Sitter spacetime does not seem to need this property. For this and other reasons, it appears that the dual system to DSSYK most naturally takes the form of a 2D gravity theory, whose physical properties are most naturally captured by dimensional reduction from 3D de Sitter gravity.

\section*{Acknowledgments}
\vspace{-1.5mm}
We thank Andreas Blommaert, Scott Collier, Jan de Boer, Juan Maldacena, Beatrix M\"uhlmann,  Erik Verlinde, Edward Witten, and Mengyang Zhang for helpful discussions and comments. The research of HV is supported by NSF grant PHY-2209997.

\appendix

\section{Relation between localized energy and the deficit angle}

In this Appendix we outline the derivation of the formula \eqref{epsi} for the energy $E(\psi)$ required to create a horizon with cosmic deficit angle $\psi$.

In flat space 3D gravity, a mass $m$ creates a conical deficit angle  $\psi = 8\pi G_N m.$ In $3$D de Sitter gravity, the notions of local mass and energy are a bit more intricate. First, spatial Cauchy slices are compact and hence there's no asymptotic boundary where one can define an ADM energy. Secondly, the geometric backreaction due the mass leads to a renormalization effect\footnote{The need for this renormalization \cite{Besken:2017fsj} can most easily be seen via the first order formulation in terms of a dreibein $e^a$ and spin connection $\omega^a$. The Einstein equations then take the form of a flatness equation of an auxiliary $SL(2,\mathbb{C})$ connection $({\cal A}^a,\bar{\cal A}^a) = (\omega^a + i e^a,\omega^a-ie^a)$. The worldline of the point particle can be formally thought of a Wilson line of this $SL(2,\mathbb{C})$ connection. It creates a delta function singularity in the curvature 2-form. This makes it necessary to regularize the definition of the Wilson line itself.} due to self-gravitational self-interaction. This effect needs to be taken into account when computing the relation between the localized energy source and the deficit angle.
It was shown in \cite{Besken:2017fsj} that the total energy $E$ of a localized mass $m$ and its self energy of contribution takes the form
\bea
\label{eself}
E = m - 2G_N m^2 .
\eea
Plugging in the relation $\psi = 8\pi G_N m$ gives the formula \eqref{epsi} 
\setcounter{equation}{5}
\bea
2\pi E = \frac{1}{16\pi G_N} (4\pi \psi - \psi^2)
\eea
for the maximal energy $E(\psi)$ contained in an SdS spacetime with cosmic deficit angle $\psi$. 
We now describe two ways to derive equations \eqref{eself}-\eqref{epsi}.

\setcounter{equation}{48}

The first derivation combines the reasonings in \cite{Besken:2017fsj} \cite{Balasubramanian:2001nb} and makes use of the Brown--York stress-energy tensor \cite{Brown:1992br}. Given a (region of) spacetime with boundary, one can compute the on-shell Einstein--Hilbert action as a function of the fixed boundary metric $h_{\mu \nu}$. The Brown--York stress-energy tensor $T_{\mu \nu}$ is then defined as 
\bea \label{brown-york stress energy tensor}
    T_{\mu \nu} = - \frac{2}{\sqrt{h}} \frac{\delta I_\text{EH}}{\delta h^{\mu \nu}}= - \frac{1}{8 \pi G_N} \left( K_{\mu \nu} - (K+1) h_{\mu \nu} \right) \, .
\eea 
Given a Killing vector $\xi^\mu$ and the normalized vector $n^\mu \equiv \xi^\mu / \sqrt{\xi^2}$, we can take a co-dimension one slice of the boundary normal to $\xi^\mu$ and define the conserved charge \cite{Brown:1992br}\cite{Balasubramanian:2001nb}
\bea \label{BYcharge}
    Q \is \int dx \, \sqrt{h_{xx}}\,  n^\mu \xi^\nu T_{\mu\nu} \, .
\eea
If the Killing vector $\xi$  corresponds to time evolution the conserved charge $Q$ indeed becomes the energy of the spacetime region under consideration. Applying this method to our setup, we should look at the behavior of the SdS metric at timelike infinity ${\cal I}^+$. The metric \eqref{esdsone} near  ${\cal I}^+$ takes the standard asymptotic de Sitter form  $ds^2 = - r^2 d\tau^2  + r^2 d\theta^2$ independent of $\psi$. We can then evaluate the Brown-York energy with respect to the Killing vector $\partial_\tau$. This Killing vector is spacelike at ${\cal I}^+$ but timelike in the static patch. The integral \eqref{BYcharge} for the SdS metric \eqref{esdsone} yields that $Q=E$.\footnote{The integral \eqref{BYcharge} for the SdS background \eqref{esdsone} looks almost identical to the Brown-York integral that compute the mass $M$ of a BTZ black hole. }

The second derivation uses the standard inductive reasoning for computing a self-energy. We start with a state with localized mass distribution with energy $E(\psi)$ inside a SdS spacetime with cosmic deficit angle $\psi$. We want to determine $E(\psi)$. The SdS metric with deficit angle $\psi$ is given by \eqref{esdsone} with $2\pi r_+ = 2\pi - \psi$. Looking in the local region near $r=0$, the euclidean SdS metric can be written as
 \bea
 \label{localds}
ds^2 = d\tau^2 + dr^2/r_+^2   +  r^2 d \varphi^2          
 \eea
with $\varphi$ and $\tau$ periodic with $2 \pi$. Note that we rescaled the euclidean time coordinate $\tau$ relative to \eqref{esdsone} so that $g_{\tau\tau} =1$. The factor of $1/r_+^2$ in the second term implies the presence of a conical singularity at $r=0$

Now let us add some extra matter so that it creates a small extra small deficit angle
 \bea
\varphi \simeq \varphi + 2\pi - \delta \psi            
 \eea
In the local metric, the localized energy source at $r=0$ that is needed to create this extra small deficit angle is
\bea
\label{deltaepsi}
\delta E = r_+ \frac{\delta\psi}{8\pi G_N}
 \eea
 The prefactor of $r_+$ is necessary to compensate for the $1/r_+^2$ prefactor in the second term in the metric \eqref{localds}.
 
The differential relation \eqref{deltaepsi} integrates to the formula \eqref{epsi} for $E(\psi)$. To complete the induction step, we again readjust the coordinates $\varphi$, $r$ so that the new metric again takes the form \eqref{localds} with $2\pi$ periodic $\tau$ and $\varphi.$ In this way, the above iterative procedure ensures that the time coordinate $\tau$ is always fixed to have the same periodicity $2\pi.$

\bibliographystyle{ssg.bst}
\bibliography{Biblio.bib}

\end{document}